\documentclass{iopart}
\usepackage{graphicx,amssymb}\usepackage{color}
\definecolor{Blue}{rgb}{0.00, 0.00, 1.00}
\definecolor{Red}{rgb}{1.00, 0.00, 0.00}

\newcommand{\nn}{\nonumber}

\newcommand{\fig}[2]{\includegraphics[width=#1]{./#2}}

\newlength{\bilderlength}

\newcommand{\E}{\mathbb{E}}
\renewcommand{\log}{\ln}
\newcommand{\eqref}[1]{(\ref{#1})}

\def\be{\begin{equation}}
\def\ee{\end{equation}}

\def\bal{\begin{align}}
\def\eal{\end{align}}

\def\bea{\begin{eqnarray}}
\def\eea{\end{eqnarray}}
\renewcommand{\log}{\ln }

\begin{document}

\title{Pickands' constant at  first order in an expansion around Brownian motion}

\author{Mathieu Delorme, Alberto Rosso and Kay J\"org Wiese}
\address{CNRS-Laboratoire de Physique Th\'eorique de l'Ecole Normale Sup\'erieure, PSL Research University,  Sorbonne Universit\'es, UPMC, 24 rue Lhomond, 75005 Paris, France.}

\begin{abstract}
In the theory of extreme values of Gaussian processes, many results are expressed in terms of the Pickands constant $\mathcal{H}_{\alpha}$. This constant depends on the local self-similarity exponent $\alpha$ of the process, i.e.\ locally it is a fractional Brownian motion (fBm) of Hurst index $H=\alpha/2$. Despite its importance, only two values of the Pickands constant are known:  ${\cal H}_1 =1$ and ${\cal H}_2=1/\sqrt{\pi}$.
Here, we extend the recent perturbative approach to fBm to include drift terms. This allows us to investigate the Pickands constant $\mathcal{H}_{\alpha}$ around standard Brownian motion ($\alpha =1$) and to derive the new exact result  $\mathcal{H}_{\alpha}=1 - (\alpha-1) \gamma_{\rm E}  + \mathcal{O}\!\left( \alpha-1\right)^{2}$.
\end{abstract}

\section{Introduction: Maximum of a Gaussian process}

The extreme-value statistics of strongly correlated variables is an active research field. However, only few general theorems for    the maximum of a set of such variables are known. Notable exceptions are  random walks \cite{Majumdar2010,BrayMajumdarSchehr2013}, the free energy of a directed polymer on a tree \cite{DerridaSpohn1988},   the eigenvalues of a random matrix \cite{TracyWidom1994}, or the extreme-values of specific Gaussian processes \cite{Sire2007,Sire2008,PiterbargBook1995,PiterbargBook2015}.

For generic Gaussian Random Processes, the tail of the distribution for large values of the maximum has been studied notably by Pickands and Piterbarg, and led to the definition  of what is now known as the {\em Pickands constant}, and which continues to be studied \cite{Harper2014,Michna2009,DebickiKisowski2008,HaanPickands1986}. 

To  appreciate  the high degree of universality of the theorems involved, we first state the original theorem of Pickands \cite{Pickands1969}, formulated for stationary processes:
Consider a stationary Gaussian process $X_t$ with mean $\E (X_t) =0$, and normalized squared variance $\label{0} {\E (X_t^2)} =1$.
By assumption,  the covariance function 
\be
r(t) := \E \left( X_{t_0} X_{t_0+t} \right) 
\ee
is independent of $t_0$. Suppose that it satisfies 
\begin{eqnarray}\label{2}
r(t) &<&1 \qquad \qquad \forall t>0\ ,\\
r(t) &\simeq& 1 -|t|^\alpha ~~~~~~\mbox{for~}t\to 0\ . \label{3}
\end{eqnarray}
Condition \eqref{2} excludes that the process is periodic, while condition  \eqref{3}  sets the scales for $X_t$ and $t$ and  defines the exponent $\alpha$. Under these circumstances, one has \cite{Pickands1969} \medskip

\noindent\underline{{\bf Theorem} (Pickands 1969):}
\begin{eqnarray}\label{5}
\mathbb{P}\!\left(\max_{t\in [0, T]}X_t>u \right) &\simeq& \Psi(u) T u^{2/\alpha}{\cal H}_\alpha \quad \mbox{as }\;u\to \infty\\
\mbox{~~~~~~~~ with }~  \Psi (u)& :=& \frac{1}{2\pi} \int_u^\infty \exp\!\left(-\frac{x^2}2\right) \rmd x \label{Psi_u}\\
\mbox{ ~~~~~~~~ and }~~~~       {\cal H}_\alpha &:=& \lim _{T\to \infty} \frac1T \E \!\left( \exp\! \left(\max\limits_{0<t<T}\chi_t \right)\right)\ . \label{Pickands-const-def}
\end{eqnarray}
The first term on the r.h.s.~of Eq.~(\ref{5}), $\Psi(u)$, is an integrated Gaussian   as expected from intuition, or more rigorously from the Borel inequality \cite{Borell1976}. The factor $T$ expresses the fact that the tail is dominated by events localized in time, and thus proportional to $T$. The non-trivial statement is that the amplitude can be calculated from  a specific process $\chi_t$ depending only on $\alpha$, and that the limit (\ref{Pickands-const-def}) exists. 

To define $\chi_t$, we first recall the definition \cite{MandelbrotVanNess1968} of a fractional Brownian motion (fBm) with Hurst exponent $H=\alpha/2$, denoted $B_t$: It is a Gaussian process starting at the origin, $B_0=0$, with mean zero, $\E (B_t)= 0$, and covariance function
\begin{equation}\label{cov_fBm}
\E \left( B_t B_s \right) = |t|^\alpha + |s|^{\alpha}-|t-s|^{\alpha}\ .
\end{equation}
The process $\chi_t$ is then defined as  a fBm with drift,
\be\label{Pick_Process}
\chi_t:=B_t-\frac12\E(B_t^2) =B_t-|t|^{\alpha}\ ,
\ee
constructed to have expectation 
$\E \left( \rme^{\chi_t}\right) = 1$.\\

Let us stress the power of this result: Apart from the Gaussian tail encoded in $\Psi(u)$, Pickands' theorem predicts not only the subleading power-law behavior $u^{2/\alpha}$, but even  (as physicists would call it) its universal amplitude ${\cal H}_\alpha$. 
\begin{figure}[b]
        \centerline{\fig{9cm}{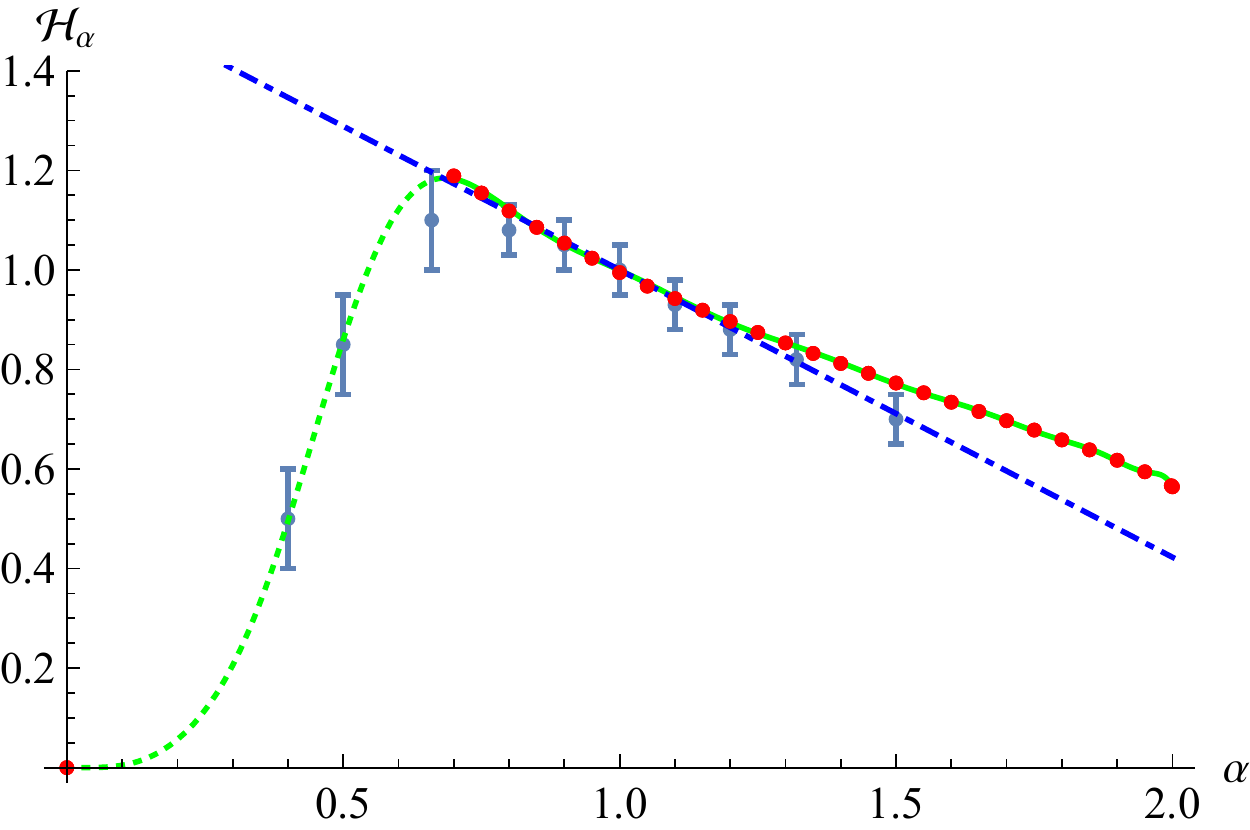}}
        \caption{Comparison of the numerical data of Ref.~\cite{DiekerYakir2014} (red dots), interpolation (green solid line), tentatively continued to $\alpha=0$ (green dashed line) and our order-$\varepsilon $ result (blue, dot-dashed). The gray-blue data points with error bars are our numerical estimates for ${\cal H}_{\alpha}$, based on Eq.\ (\ref{PiterbargTh}), see \ref{app:bridge2Pickands} for details.}
        \label{f:Dieker-data}
\end{figure}

A major challenge remains, namely  evaluation of Pickands' constant. Only the cases where fBm reduces to  standard Brownian motion ($\alpha=1$),  and where fBm is an affine process ($\alpha=2$, i.e.\ a straight line) are known,
\be
{\cal H}_{1} = 1 \qquad \textrm{ and } \qquad
{\cal H}_{2} = \frac1{\sqrt \pi}\ .
\ee
There is yet no analytical result for other values of $\alpha$. In this letter, we use a path integral formulation, evaluated perturbativly around Brownian motion, to  show that 
\be\label{us-Pickands}
{\cal H}_{\alpha} = 1 - \gamma_{\rm E}(\alpha-1) + {\cal O}(\alpha-1)^2\ ,
\ee
where $\gamma_{\rm E}$ is Euler's constant. 

For other values of $\alpha$, only numerical estimations exist, see figure \ref{f:Dieker-data}. These are difficult, since  statistical fluctuations in Pickands' definition (\ref{Pickands-const-def}) are large. E.g.\ for a Brownian the estimator (\ref{Pickands-const-def}) convergences as $1/\sqrt{T}$ \cite{DiekerYakir2014}. A representation with a much better   convergence has been given by Dieker and Yakir \cite{DiekerYakir2014}:
\medskip

\leftline{\underline{{\bf Theorem} (Dieker and Yakir, 2014):}}
\be\label{DiekerYakir2014-estimator}
{\cal H}_\alpha = \lim_{T\to \infty} \E \!\left( \frac{\rme^{\max_{0<t<T}\chi_t }}{\int_0^T \rme^{\chi_t } \rmd t}\right)\ .
\ee
Thus effectively the inverse of $T = \E \left(  \int_0^T  \rme^{\chi_t}\, \rmd t  \right)$ can be moved inside the expectation value $\E \left( \rme^{\max_{0<t<T}\chi_t }\right)$. 
The estimator (\ref{DiekerYakir2014-estimator}) converges much better  than Pickands original one, leading to the results presented on figure \ref{f:Dieker-data} (red dots).\\

Let us conclude this introduction by another remarkable theorem due to V.I.~Piterbarg \cite{PiterbargBook1995,PiterbargBook2015}, which extends  Pickands' theorem by relaxing the stationarity hypothesis.
Suppose that a random process $X_t$ with zero mean is defined on the interval $[0,T]$, and has a unique time $t_0 \in (0,T)$ of maximal variance, normalized to $1$. Further suppose that for some positive $a$, $c$, $\alpha$ and $\beta$ the variance and covariance functions satisfy
\begin{eqnarray}\label{12}
~~\sigma(t) &:=& \sqrt{\E(X_t^2)} \;= 1 - a |t-t_0|^\beta  \qquad~~ \mbox {for }t\to t_0\ ,\\
r(t,s) &:=& \E\left( X_t X_s \right) = 1- c |t-s|^\alpha 
  \quad \mbox{for }~ t\to t_0 \mbox{ and } s \to t_0\ . \label{13}
\end{eqnarray}
One finally needs a weak regularity condition, namely that for some $\gamma$ and $G$ positive, 
$\E \left(X_t-X_s \right) ^2 \le G |t-s|^{\gamma}$. 
The theorem D.3 of Piterbarg \cite{PiterbargBook1995} (see also \cite{PiterbargBook2015}) then distinguishes several cases. We only state the one which is relevant below.\medskip

\leftline{\underline{{\bf Theorem} (Piterbarg 1978):}}\smallskip

If $\beta>\alpha$, then 
\be\label{PiterbargTh}
\fl \qquad \mathbb{P} \!\left(\max\limits_{t\in[0,T]} X_t>u\right) = \frac{2 {\cal {\cal H}_\alpha } \Gamma(1+1/\beta) c^{\frac 1 \alpha}}{ a^{\frac 1 \beta}} u^{\frac{2}{\alpha} -\frac{2}{\beta}} \Psi(u)~~\mbox{ as  } u \to \infty \ ,
\ee
with ${\cal H}_{\alpha}$ and $\Psi(u)$ as defined in Eqs.~\eqref{Psi_u} and \eqref{Pickands-const-def}. 

This beautiful theorem   applies to a  fractional Brownian bridge defined on $[0,1]$ and reproduces the Pickands constant of Eq.~(\ref{us-Pickands}), see   \ref{app:bridge2Pickands}.\\

To simplify the discussion in the next sections, we introduce a process $z_t$ with an arbitrary drift strenght $\mu$
\begin{equation}\label{zDef}
z_t=B_t+\mu |t|^{\alpha}\ .
\end{equation}
Setting $\mu = -1$ allows us to recover $z_t=\chi_t$, as defined in Eq.~\eqref{Pick_Process}.  Pickands' constant can also be computed by setting $\mu=1$, using
\begin{equation}\label{PickandsBis}
\mathcal{H}_{\alpha} = \lim\limits_{T \to \infty} \frac1T \E(\rme^{-\min_{t\in[0,T]}z_t})\ .
\end{equation}

\section{Brownian with drift, and its Pickands constant ($\alpha=1$)}        
We recall   some results about Brownian motion with drift which are useful to expand Pickands' constant around $\alpha=1$. 

For $\alpha =1$,  the fBm process $B_t$ is  a standard Brownian motion, with covariance $\E(B_t B_s)= 2 D \min(t,s)$, and diffusion constant $D$.
The propagator $P^+_{\mu}$ of the process $z_t$ defined in Eq.~\eqref{zDef}, with positivity constraint is  \footnote{This result, obtained by the method of images, is easily checked to  satisfy the diffusion equation with the appropriate boundary conditions.}
\begin{eqnarray}\label{PropagatorBrownian}
P^+_{\mu}(x_0,x,T)&:=&\,\partial_x \mathbb{P}\!\left(z_T < x, \min\limits_{t\in[0,T]} z_t >0\, \Big| z_0 = x_0 \right) \nn \\
&=&\, \frac{\rme^{ \frac{\mu}{2D}(x-x_0) - \frac{\mu^2}{4D}T}}{\sqrt{4 \pi D T}}\!\left(\rme^{- \frac{(x-x_0)^2}{4D T}}-e^{- \frac{(x+x_0)^2}{4D T}}\right)\nn \\
&=&\,\rme^{ \frac{\mu}{2D }(x-x_0) - \frac{\mu^2}{4D}T} P_0^+(x_0,x,T)\ .
\end{eqnarray}
Here $P_0^+$ is the  propagator for the process without drift, i.e.\ $\mu = 0$. To compute Pickands' constant we choose $\mu=D=1$, cf.\ Eq.~\eqref{PickandsBis}. We can recover     a generic diffusion constant $D$ (with $\mu=D$), by setting $T\to D T$ as can be checked on Eq.\ \eqref{PropagatorBrownian}. The survival probability $Q$ of this process, which is defined as the probability to remain positive up to time $T$ while starting at $x_0>0$, can be computed from $P_{\mu}^+$ as
\begin{eqnarray}\label{SurvivalBrownian}
\fl Q_{\alpha=1}(x_0,T)= \int_0^{\infty} \!\!\!\rmd x\,P^+_{\mu=1}(x_0,x,T) = \frac12\! \left[\mbox{erf}\!\left(\frac{x_0+T }{2 \sqrt{T}}\right)-\rme^{-x_0 } \mbox{erfc}\!\left(\frac{x_0-T }{2 \sqrt{T}}\right)+1\right]. \nn\\
\end{eqnarray}  
From this result we can extract the distribution of $m$, defined as $m:=-\min_{t\in[0,T]}z_t$, 
\begin{equation}\label{MaxDistribBrownian}
\fl \qquad \mathcal{P}^T_{\alpha=1}(m)= \partial_{m}Q_{\alpha=1}(m,T)=\frac{1}{2} \rme^{-m} \mbox{erfc}\left(\frac{m-T}{2 \sqrt{T}}\right)+\frac{\rme^{-\frac{(m+T)^2}{4 T}}}{\sqrt{\pi T} }\ .
\end{equation}
The result \eqref{MaxDistribBrownian} allows us to  compute Pickands' constant via its main definition \eqref{PickandsBis}:
\begin{eqnarray}\label{9}
\fl && \int_0^{\infty}\!\!\!\rmd m \, e^m \mathcal{P}^T_{\alpha=1}(m) = \left(\frac{T}{2}+1\right)\!\left[\mathrm{erf}\!\left(\frac{\sqrt{T}}{2}\right)+1\right] + \sqrt{\frac{T}{\pi}}e^{-\frac{T}{4}} \nn\\ 
&&\hphantom{\int_0^{\infty}\!\!\!\rmd m \, e^m \mathcal{P}^T_{\alpha=1}(m)}\!\! \stackrel{T \to \infty}{\simeq} T +2  + \mathcal{O}(e^{-\frac{T}{4}}).  
\end{eqnarray}
The Pickands constant is the coefficient of the linear term in the large-$T$ asymptotics of Eq.~\eqref{9}. We thus recover the known result for the Brownian, $\mathcal{H}_{1} =1$.

\section{Perturbative expansion around  Brownian motion: $\alpha=1 + 2\varepsilon $}
\subsection{Action}
For \be
\alpha=1+2\varepsilon
\ee
 with $\varepsilon$ a small parameter, we construct in   \ref{A:action} the action for the process $z_t$, defined in Eq.~\eqref{zDef} with $\mu=1$. This follows the ideas of \cite{DelormeWiese2016b,WieseMajumdarRosso2010,DelormeWiese2015,DelormeWiese2016}. One writes
\begin{equation}\label{ActionExpansion}
\mathcal{S}[z_t]=\mathcal{S}_0[z_t]+ \varepsilon\,  \mathcal{S}_{1}[z_t]+\mathcal{O}(\varepsilon^2)\ ,
\end{equation}
with
\begin{eqnarray}\label{Action}
\fl\qquad \mathcal{S}_0[z_t]=&  \int_{0}^{T}\rmd t\, \frac{\dot{z}_t^2}{4 D_{\varepsilon ,\tau}}  - \frac{ (z_T-z_0)}{2}+ \frac{D_{\varepsilon ,\tau} T }{4}\ ,  \\
\fl\qquad \mathcal{S}_1[z_t]=&-  \frac12 \int_{0}^{T}\rmd t\,\dot{z}_t \log\!\left(\frac{t}{T-t}\right)- \frac12 \int_{0}^{T-\tau}\!\!\!\rmd t_1\int_{t_1+\tau}^{T}\!\!\rmd t_2\, \frac{\dot{z}_{t_1}\dot{z}_{t_2}}{t_2-t_1}\ .~~~~~~~~
\label{Action1}
\end{eqnarray}
We recognise $\mathcal{S}_0$ as the standard Brownian action with a  diffusion constant  \cite{DelormeWiese2016}
\begin{equation}\label{DiffConstant}
D_{\varepsilon ,\tau}=1+ 2\varepsilon [1+\ln(\tau)] + \mathcal{O}(\varepsilon^2)\ ,
\end{equation}
and a linear drift $\mu=D_{\varepsilon,\tau}$. The time $\tau$ is a regularization cutoff for coinciding times (an
UV cutoff), necessary to define   perturbation theory. It has no impact on the distribution of observables which can be extracted from the path integral \cite{WieseMajumdarRosso2010,DelormeWiese2016}.

\subsection{Pickands' constant}
To investigate Pickands' constant, we start with a path-integral representation for the survival probability of the process $z_t$, an idea introduced in Refs.~\cite{MajumdarSire1996,SireMajumdarRudinger2000},
and   developed for the situation at hand in Refs.~\cite{WieseMajumdarRosso2010,DelormeWiese2016b,DelormeWiese2016,DelormeWiese2015}:
\begin{equation}\label{SurvivalPathIntegral}
Q_{\alpha}(m,T)= \frac{1}{Z^{\rm N}(T)}\int_{0}^{\infty} \rmd x \int_{z_0=m}^{z_T=x} \mathcal{D}[z_t] \,\Theta[z_t]\rme^{-\mathcal{S}[z_t]}\ ,
\end{equation}
where $\Theta[z_t]$ constrains the path $z_t$ to remain positive; the normalisation constant $Z^{\rm N}(T)$ is the sum over all paths without the constraint $z_{0}=m$ (and thus independent of $m$). Computing the path integral in Eq.~\eqref{SurvivalPathIntegral} within the $\varepsilon$-expansion of the action \eqref{ActionExpansion} allows us to write
\begin{eqnarray}\label{expansion}
\fl Z^{\rm N}(T)Q_{\alpha}(m,T)&=&Z_0^+(m,T)+ \varepsilon  Z_1^+(m,T) + \mathcal{O}(\varepsilon ^2)\\
\fl &=&\langle\Theta[z_t]\rangle_0 + \varepsilon\! \left[\langle\Theta[z_t]\mathcal{S}_1[z_t]\rangle_0+2(1+\ln \tau)T\partial_TZ_0^+(m,T)\right] + \mathcal{O}(\varepsilon^2)\nn \ . \label{expansion2}
\end{eqnarray}
The symbol $\langle... \rangle_0$ denotes averages over paths $z_t$ with  the standard Brownian action with drift ($\mu=D=1$), initial conditon $z_0=m$ and a free end-point $z_T$. Thus, the zeroth-order term \be
Z_0^+(m,T)\equiv\langle\Theta[z]\rangle_0 \equiv Q_{\alpha=1}(m,T)
\ee  is the survival distribution of the Brownian  as given in Eq.\ \eqref{SurvivalBrownian}. For the order-$\varepsilon $ term $Z_1^+$, there is a contribution due to the non-local correction of the action $\mathcal{S}_1$, cf. Eq.~\eqref{Action1}, and a contribution due to the rescaling of the diffusive constant (and the drift) in $\mathcal{S}_0$, $D=1\to D_{\varepsilon ,\tau}$.

Before expliciting these terms, we show how this leads to the Pickands constant. Using $Z_N(T)= \lim_{m \to \infty} Z_N(T) Q_{\alpha}(m,T)$, (note that $Q_{\alpha}$ is the  cumulative distribution), we arrive at
\begin{equation}
\fl \qquad Q_{\alpha}(m,T)=Z_0^+(m,T)[1-\varepsilon  \lim\limits_{m\to \infty}Z_1^+(m,T) ]+ \varepsilon  Z_1^+(m,T)+\mathcal{O}(\varepsilon ^2)\ .
\end{equation}
As for   $\alpha=1$ in Eq.~\eqref{9}, the Pickands constant is obtained from the large-$T$ asymptotics of
\begin{eqnarray}\label{PickandsExpansion}
\fl &&\int_{0}^{\infty} \rmd m\, e^m \partial_m Q_{\alpha}(m,T) = \int_{0}^{\infty} \rmd m\, e^m \partial_m Z_0^+(m,T)\\
\fl && +\,\varepsilon  \left[\int_{0}^{\infty} \rmd m\, e^m \partial_m Z_1^+(m,T) -\lim\limits_{m \to \infty}Z_1^+(m,T) \int_{0}^{\infty} \rmd m\, e^m \partial_m Z_0^+(m,T) \right]+\mathcal{O}(\varepsilon ^2)\nn\ .
\end{eqnarray}
The first term was already computed in Eq.\ \eqref{9}. For the order-$\varepsilon $ term, the function $Z_1^+(m,T)$ can be expressed from the bare propagator $P_{\mu=1}^+$, given in Eq.\ \eqref{PropagatorBrownian}, and its cumbersome Laplace transform $\tilde Z_1^+(m,s)$   derived in \ref{A:details}. The asymptotics
\begin{eqnarray}\label{Asymptotic1}
\fl \int_0^{\infty}\!\!\!\rmd m\,e^{m} \partial_m Z_1^+(m,T) \stackrel{T\to \infty}{=} \frac{T^2}{2}\! \left[\ln\!\left(\frac{T}{\tau}\right)-1\right]+ {T} \!\left[\ln\!\left(\frac{T}{\tau}\right)-1-2 \gamma_{\rm E} \right]+ \mathcal{O}\big(\ln(T)\big) \nn\\
\end{eqnarray}
and
 \begin{equation}\label{AsymptoticNormalisation}
 \lim\limits_{m \to \infty} Z_1^+(m,T)=  \frac{T}{2} \left[\ln\!\left(\frac{T}{\tau}\right)-1\right]\ ,
\end{equation}
allows us to compute Pickands' constant at order $\varepsilon $. Combining these contributions according to Eq.~\eqref{PickandsExpansion} cancels the $\tau$ dependence, as it should, and finally gives
\begin{equation}
\mathcal{H}_{1+2\varepsilon }=1-2\varepsilon  \gamma_{\rm E}+ \mathcal{O}(\varepsilon ^2) \equiv 1 - (\alpha-1) \gamma_{\rm E}  + \mathcal{O}\!\left( \alpha-1\right)^{2}\ ,
\end{equation}
where $\gamma_{\rm E}$ is the Euler-Mascheroni constant, whose numerical value is $\gamma_{\rm E}\approx 0.577$. 

This result, which gives the derivative of the Pickands constant at $\alpha =1$,   compares favourably  to the extensive numerical simulations of Ref.~\cite{DiekerYakir2014}  plotted on figure \ref{f:Dieker-data}. Though much less precise, it is also in agreement with our results obtained by numerical simulations of the maximum of a fBm bridge, using Eqs.~(\ref{PiterbargTh})--(\ref{PiterbargTh2}).

\subsection{Distribution of $m$ at large $T$}

\begin{figure}[t]
\centerline{\includegraphics[width=9cm]{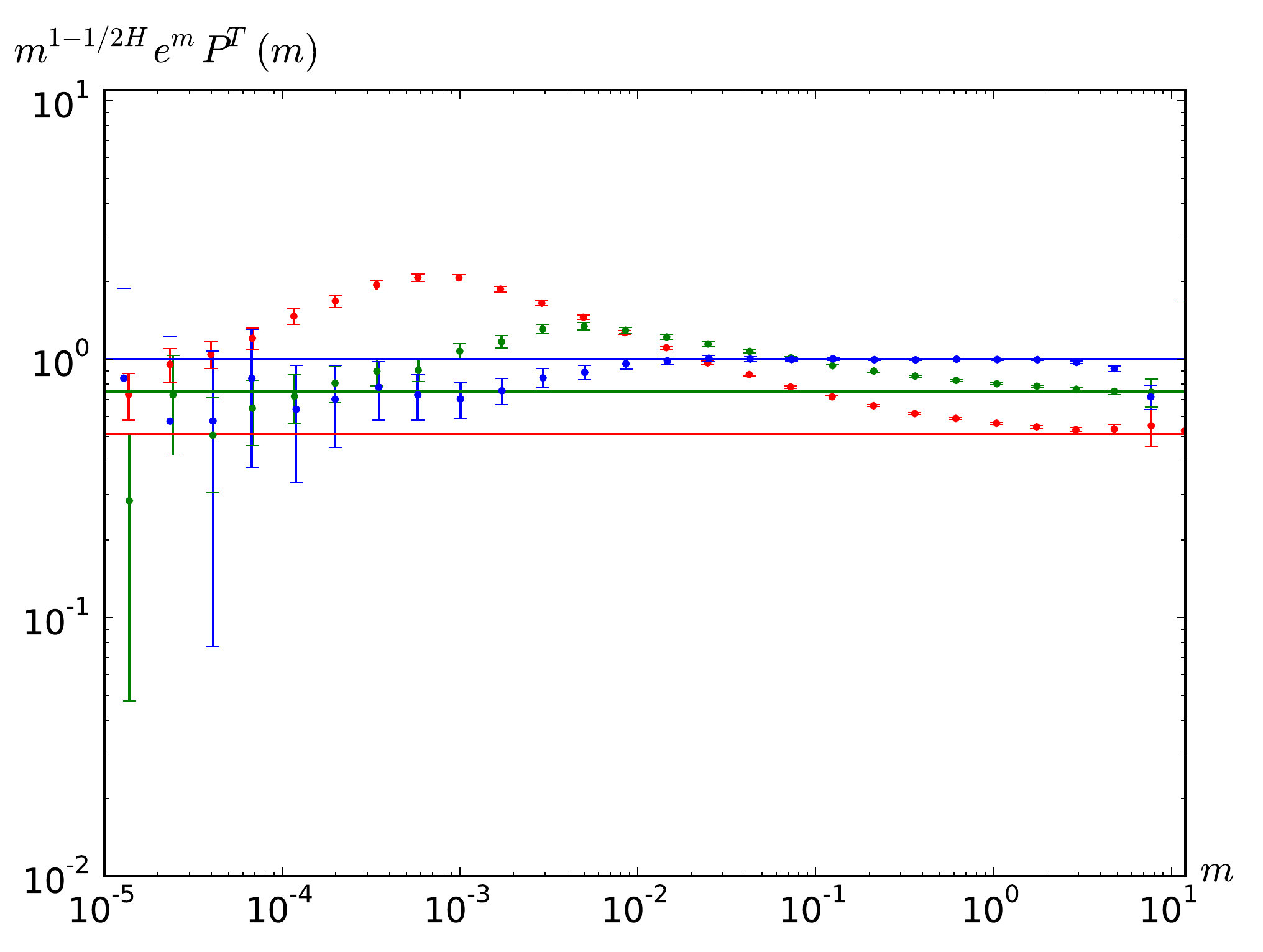}}
\caption{Test on the asymptotic behavior of $\mathcal{P}_{\alpha}^{\infty}(m)$, for $\alpha=1.$ (blue), $\alpha=1.2$ (green) and $\alpha=1.5$ (red). Plain lines represent the conjectured limits for large $m$, using the numerical value of $\mathcal{H}_{\alpha}$ from \cite{DiekerYakir2014}. Simulation parameters are $T=8$, $dt=2^{-14}$, with $10^6$ samples.}
\label{f:our-conjecture-test}
\end{figure}

For standard Brownian motion,  $\alpha=1$, the distribution $\mathcal{P}_{\alpha=1}^{T}(m)$ given in Eq.~\eqref{MaxDistribBrownian} has the interesting property to converge to a non-trivial limit when $T \to \infty$, namely
\begin{equation}
\mathcal{P}_{\alpha=1}^{ \infty}(m) = \lim\limits_{T \to \infty} \partial_m Q_{\alpha =1}(m,T)= \rme^{-m}\ .
\end{equation}
Using the same expansion as in Eq.~\eqref{PickandsExpansion}, we can express this distribution for $\alpha=1+2\varepsilon $,
\begin{eqnarray}
\fl \mathcal{P}_{\alpha}^T(m) = \partial_m Z_0^+(m,T)
+\,\varepsilon \left[ \partial_m Z_1^+(m,T) -\Big(\lim\limits_{m \to \infty}Z_1^+(m,T) \Big) \partial_m Z_0^+(m,T) \right]+\mathcal{O}(\varepsilon ^2)\nn\ .\\
\end{eqnarray}
The expression of $\tilde{Z}_1^+(m,s)$  given in \ref{A:details} encodes $\mathcal{P}_{\alpha}^T(m)$ for a generic $T$, but we   restrict ourselves to the large-$T$ limit for simplicity. Using the asymptotics
\begin{eqnarray}\label{Asymptotic2}
\fl \partial_m Z_1^+(m,T) \stackrel{T \to \infty}{=}-2 \rme^{-m}\left\{1+\gamma_{\rm E}+\ln(m)+ \frac{T}{4}\!\left[1+\log\!\left(\frac{\tau}{T}\right)\right]\right\}- 2\mbox{Ei}(-m)+\mathcal{O}\Big(\frac1T\Big),\nn\\
\end{eqnarray}
and the one given in Eq.~\eqref{AsymptoticNormalisation}, we see that $\mathcal{P}_{\alpha}^T(m)$ converges at large $T$ to a non-trivial distribution, 
\begin{equation}
\fl \qquad \mathcal{P}_{\alpha=1+2\varepsilon }^{\infty}(m)=\rme^{-m} \Big\{1-2 \varepsilon  \big[1+\gamma_{\rm E} + e^m \mbox{Ei}(-m)+\log(m)\big]\Big\}+\mathcal{O}(\varepsilon ^2)\ .
\end{equation}
This is in agreement with the following conjecture: For all $\alpha \in (0,2)$, the distribution $\mathcal{P}_{\alpha}^T(m)$ converges to a distribution $\mathcal{P}_{\alpha}^{\infty}$(m) which has the   large-$m$ asymptotics
\begin{equation}\label{conjecture}
\mathcal{P}_{\alpha}^{\infty}(m) \stackrel{m \to \infty}{\simeq} \frac{\mathcal{H}_{\alpha}}{\alpha} m^{\frac{1}{\alpha}-1}\rme^{-m}\ .
\end{equation}
This conjecture is numerically tested on figure \ref{f:our-conjecture-test}. It can be motivated heuristically, see \ref{A:conjecture}.

\section{Conclusions}
In this letter, we   derived the linear term in the expansion of the Pickands constant around Brownian motion. Apart from the Pickands constant at $\alpha=1$ and $\alpha=2$, this is the only analytically available information we have today. 

It would be interesting to continue this approach to higher orders. 
While the quadratic term seems feasible, it is rather difficult to evaluate, and has to be left for future research. 

As our methods allow us to obtain the full distribution of the maximum, and not only its limiting behavior for large arguments, other questions can be posed. A particularly interesting one is the probability distribution of the maximum of a fBm with an unconstraint endpoint. From \cite{PiterbargBook1995,PiterbargBook2015} we know that at $\alpha=1$ this behavior changes. For $\alpha<1$ it  is non-trivial as in Eq.~(\ref{PiterbargTh}),  while for $\alpha>1$ the tail is simply given by the distribution at the endpoint. For $\alpha$ close to 1 both terms will contribute in a non-trivial way yet to be determined.

\section*{Aaknowledgments}
We thank V.I.~Piterbarg, S.\ Kobelkov, S.N.\ Majumdar, K.\ Mallick and W.\ Werner for helpful  discussions. 
Part of this work has been supported by PSL through grant
ANR-10-IDEX-0001-02-PSL.

\appendix
\section{Derivation of the action in presence of a drift}
\label{A:action}
Here we derive  the action for the process $z_t = B_t+ \mu \,t^{\alpha}$, where $X_t$ is a fractional Brownain motion, close to standard Brownian motion, i.e.\ with parameter $\alpha=1+2\varepsilon $, and $\varepsilon $ small. As the process $B_t$ is Gaussian, its action is given by its covariance function $G^{-1}(t_1,t_2)=\E ( B_{t_1} B_{t_2} )$,
\begin{equation}
\mathcal{S}[B]=\frac{1}{2}\int_{t_1,t_2} \dot{B}_{t_1} G(t_1,t_2)\dot{B}_{t_2}\ .
\end{equation}
While it is not possible to derive a simple closed  expression for a generic value of $\alpha$,   we can express the action $\mathcal{S}$ in an $\varepsilon $-expansion. This was done in Ref.\ \cite{WieseMajumdarRosso2010} for the process without drift; the result reads
\begin{equation}
\mathcal{S}[B]=  \int_{0}^{T}\! \!\rmd t\, \frac{\dot{B}_t^2}{4 D_{\varepsilon ,\tau}}  -  \frac{\varepsilon }{2} \int_{0}^{T-\tau}\!\!\rmd t_1\int_{t_1+\tau}^{T}\!\!\rmd t_2\, \frac{\dot{B}_{t_1}\dot{B}_{t_2}}{t_2-t_1} + \mathcal{O}(\varepsilon ^2)\ .
\end{equation}
The first term involves a rescaled diffusion constant $D_{\varepsilon ,\tau}=1+2\varepsilon (1+\ln \tau)+\mathcal{O}(\varepsilon^2)$. From this, it is possible to obtain the action for $z_t$ by     changing  variables $\dot{B}_t \to \dot{z}_t - \mu \left[1+2\varepsilon (1+\ln t)\right]+\mathcal{O}(\varepsilon^2)$.
Expanding each term of the action, we get
\begin{eqnarray}
\fl \int_{0}^{T}\! \!\rmd t\, \frac{\dot{B}_t^2}{4 D_{\varepsilon ,\tau}} &\to & \int_{0}^{T}\! \!\rmd t\, \frac{\dot{z}_t^2}{4 D_{\varepsilon ,\tau}}- \mu \frac{z_T-z_0}{2 D_{\varepsilon ,\tau}} +  \frac{\mu^2 T}{4 D_{\varepsilon ,\tau}} \nn \\
\fl && - \varepsilon  { \mu} \int_{0}^{T}\!\! \rmd t\,\dot{z}_t(1+\ln t) + \varepsilon    {\mu^2 T\ln(T)} +\mathcal{O}(\varepsilon^2) \ ,
\end{eqnarray}
and
\begin{eqnarray}
\fl \int_{0}^{T-\tau}\!\!\rmd t_1\int_{t_1+\tau}^{T}\! \!\rmd t_2\, \frac{\dot{B}_{t_1}\dot{B}_{t_2}}{t_2-t_1} &\to&  \int_{0}^{T-\tau}\!\!\rmd t_1\int_{t_1+\tau}^{T}\!\!\rmd t_2\, \frac{\dot{z}_{t_1}\dot{z}_{t_2}}{t_2 -t_1} - \mu^2 T\left[\ln\!\left(\frac{\tau}{T}\right)+1\right]\nn\\
\fl &&- \mu \int_0^T \rmd t\, \dot{z}_t\left[ \log\!\left(\frac{t}{\tau}\right)+\log\!\left(\frac{T-t}{\tau}\right) \right]+\mathcal{O}(\varepsilon )\ .
\end{eqnarray}
There are some simplifications:
\begin{equation}
\fl \mu ^2\frac{ T}{4 D_{\varepsilon,\tau}} + \varepsilon  {\mu^2 T \log(T)}+ \varepsilon \frac{\mu^2 T}{2}\left[\ln\!\left(\frac{\tau}{T}\right)+1\right]= \mu^2 \frac{T^{1+ 2 \varepsilon }}{4} +\mathcal{O}(\varepsilon ^2)
\end{equation}
\begin{eqnarray}
\fl \int_0^T \!\!\!\!\rmd t\, \dot{z}_t \log\!\left(\frac{t(T-t)}{\tau^2}\right)-  2\int_{0}^{T}\!\!\!\!\rmd t\,\dot{z}_t (1+\ln t) = \int_0^T \!\!\!\!\rmd t\, \dot{z}_t\log\! \left(\frac{T-t}{t}\right) - 2(z_T-z_0) (1+\log\tau) \nn\\
\end{eqnarray}
After recombining  these terms, we  obtain the rather compact expression
\begin{eqnarray}\label{ActionAppendix}
\fl \mathcal{S}[z_t]=&  \int_{0}^{T}\rmd t\, \frac{\dot{z}_t^2}{4 D_{\varepsilon ,\tau}}  - \mu\frac{ z_T-z_0}{2}+ \mu^2\frac{ T^{1+2\varepsilon } }{4} \nn \\
\fl &-\mu \frac{\varepsilon }{2} \int_{0}^{T}\rmd t\,\dot{z}_t \log\!\left(\frac{t}{T-t}\right)- \frac{\varepsilon }{2}  \int_{0}^{T-\tau}\!\!\!\rmd t_1\int_{t_1+\tau}^{T}\!\!\rmd t_2\, \frac{\dot{z}_{t_1}\dot{z}_{t_2}}{t_2-t_1}+ \mathcal{O}(\varepsilon ^2)\ .
\end{eqnarray}
The last term of the first line does not depend on $z_t$, thus acts   as a global normalisation which has no impact on the observables we compute from this action. We choose to change it to $\mu^2 T D_{\varepsilon ,\tau}/4$ for simplicity and fix $\mu=1$, which finally gives the expressions \eqref{ActionExpansion} and \eqref{Action} of the main text.

\section{Details of the calculations}
\label{A:details}
In this appendix, we give the details of the computation for the order-$\varepsilon $ correction in the path integral \eqref{SurvivalPathIntegral}. The difficult contribution in Eq.\ \eqref{expansion} is $Z_{1,AB}^+:=\langle\mathcal{S}_1[z_t] \Theta[z_t]\rangle_0$, which we now decompose into two terms $Z_{1,AB}^+=Z_{1A}^++Z_{1B}^+$ using the expression of $\mathcal{S}_1$ given in the action \eqref{Action}:
\begin{equation}\label{Z1A}
Z^+_{1A}(m,T) =\frac{1}{2} \int_{0}^{T-\tau}\rmd t_1\int_{t_1+\tau}^{T} \rmd t_2 \frac{\left\langle\dot{z}_{t_1}\dot{z}_{t_2} \Theta[z] e^{ \frac{Z_T-Z_0}{2} - \frac{T}{4}}\right\rangle_0}{t_2-t_1}\ ,
\end{equation}
and
\begin{equation}\label{Z1B}
Z^+_{1B}(m,T) = \frac{1}{2}\int_{0}^{t} \rmd t \left\langle\dot{z}_t\, \Theta[z] e^{\frac{Z_T-Z_0}{2} -  \frac{T}{4}}\right\rangle_0 \log \left( \frac{t}{T-t}\right)\ .
\end{equation}
The averages $\langle... \rangle_0$ denote averages with respect to the standard Brownian action, with no drift, as the drift is now enforced   by the exponential factors. We can express these averages in terms of the drift-free bare propagator  with  positivity constraint,   $P_0^+$. Following the diagrammatic rules defined in Ref.~\cite{DelormeWieseUnPublished}, the first correction can be written after a   Laplace transform $T \to s$ as
\begin{equation}\label{Z1ALaplace}
\fl \tilde Z_{1A}^+(m,s)
= 2\int_{x_i,y>0}\!\!\!e^{\frac{x_3-m}{2}}\tilde P_0^+\!\left(m,x_1,\bar s\right)\partial_{x_1} \tilde P_0^+\!\left(x_1,x_2,\bar s+y\right)\partial_{x_2} \tilde P_0^+\!\left(x_2,x_3,\bar s\right)\ .
\end{equation}
We   introduced $\bar s := s + 1/4$, a shifted Laplace variable due to the term $e^{- T /4}$. As explained in Ref.~\cite{WieseMajumdarRosso2010}, each $\dot{z}_{t_i}$ in \eqref{Z1A} corresponds to a factor of $  2\partial_{x_i}$ acting to the following propagator in Eq.~\eqref{Z1ALaplace}. To account for the factor of $(t_2-t_1)^{-1}$, we use the identity $(t_2-t_1)^{-1} = \int_{y>0}e^{-y(t_2-t_1)}$ which produces a shift in the second propagator by a new variable $y$ wkich we need to integrate over. We recall the expression of the propagator in Laplace variables,  
\begin{equation}
\tilde P_0^+(x_1,x_2,s) =  \frac{e^{-\sqrt{s}|x_1-x_2|}-e^{-\sqrt{s}(x_1+x_2)}}{2\sqrt{s}}\ .
\end{equation}
The second correction, due to the non linearity in the drift, is given by  
\begin{eqnarray}\label{Z1Bbis}
\fl Z^+_{1B}(m,T)=   \int\limits_{x_i>0}\!\!\int_0^T\!\!\!\rmd t\, e^{\frac{x_2-m}{2}} P_0^+(m,x_1,t) \partial_{x_1} P_0^+(x_1,x_2,T-t) \log\!\left( \frac t {T-t}\right) e^{-\frac{T}{4}}\ . \nn\\
\end{eqnarray}
In order to compute its Laplace transform, we use the   integral representation\begin{equation}
\ln\!\left(\frac t{T-t}\right) = \int_{0}^{\infty} \frac{\rmd y}{y} \left(e^{-y (T-t)}-e^{-y t}\right)\ .
\end{equation}
Inserting this into  Eq.~\eqref{Z1Bbis} and taking the Laplace transform gives  
\begin{eqnarray}\label{Z1BLaplace}
\nn
\fl \tilde Z_{1B}^+(m,s) &=&  \int_{0}^{\infty} \frac{\rmd y}{y} \int_{x_i>0} e^{\frac{x_2-m}{2}} \Big[ \tilde P_0^+(m,x_1,\bar s)\, \partial_{x_1} \tilde P_0^+(x_1,x_2,\bar s+y)     \\
\fl && ~~~~~~~~~~~~~~~~~~~~~~~~~~~~- \tilde P_0^+(m,x_1,\bar s+y)\, \partial_{x_1} \tilde P_0^+(x_1,x_2,\bar s)\Big]\ .
\end{eqnarray}
For both $\tilde{Z}_{1A}^+$ and $\tilde Z_{1B}^+$, the integrals over the space variables $x_i$ can be computed quite easily, as the Laplace-transformed propagator $\tilde{P}_0^+$ is exponential in these variables (contrary to  the time-dependent propagators, where the dependence is Gaussian). For the integral over $y$, $\tilde{Z}_{1A}^+$ has a logarithmic divergence at large $y$ which corresponds to the UV divergence when $t_2 \to t_1$ in Eq.~\eqref{Z1A}. The necessary large-$y$ cutoff $\Lambda$ (such that the integration over $y$ is performed in the interval $[0,\Lambda]$) equivalent to the UV cutoff $\tau$ is given by $\Lambda = e^{-\gamma_{\rm E}}/\tau$,  \footnote{As explained in Ref.~\cite{DelormeWieseUnPublished},  this comes from the requirement: $\int_0^T \rmd t \int_0^{\Lambda} e^{-yt} = \log(\Lambda T)+\gamma_{\rm E}+\mathcal{O}(e^{-T\Lambda}) \stackrel{!}{=} \ln(T/\tau)=\int_{\tau}^{T} \frac{\rmd t}{t}$.}. 

Combining these two terms finally gives (remind  $\bar s=s+\frac14$)
\begin{eqnarray}
\fl \lefteqn{s^2\Big[ \tilde Z_{1A}(m,s)+\tilde Z_{1B}(m,s)\Big]} \nn \\
\fl & =& -\frac{e^{\left(\sqrt{\bar s}-\frac{1}{2}\right) m}}{8 \sqrt{\bar s}} \Big[8 \bar s^{3/2} (m+1)+8s \bar sm+4\bar s-2 \sqrt{\bar s} (m-1)-1\Big] \mbox{Ei}\!\left(-2 \sqrt{\bar s} m\right)\nn\\
\fl && +\frac{e^{-\left( \sqrt{\bar s}+\frac{1}{2}\right) m}}{16\sqrt{\bar s}} \bigg[\left(8 \bar s^{3/2}+8s \bar s m+4 \bar s+2 \sqrt{\bar s}-1\right) \big( \log (4 \bar s \tau) +1 + \gamma_{\rm E} \big)-8\sqrt{\bar s}\nn\\
\fl &&~~~~~~~~~~~~~~~~~~~~+\left(8 \bar s^{3/2}-8s\bar sm +4\bar s-6 \sqrt{\bar s}-1\right) \left(\log\!\left(\frac{m^2}{\tau}\right)-1 + \gamma_{\rm E}\right)\bigg]\nn\\
\fl&& +\frac12 \Big[\mbox{Ei}\!\left(-\frac{m}{2}-m \sqrt{\bar s}\right)+e^{-m} \mbox{Ei}\!\left(\frac{m}{2}- m\sqrt{\bar s}\right)-\log (s \tau)-\gamma_{\rm E}\Big]
\ .\end{eqnarray}
From this expression, and denoting $\tilde{Z}_{1AB}^+:=\tilde{Z}_{1A}^++\tilde{Z}_{1B}^+$ , it is possible to compute the asymptotics used in the main text, first in terms of the Laplace variable:   
\begin{eqnarray}
\fl \tilde Z_{1AB}(m,s) &\stackrel{s \to 0}{\simeq} &\frac{(e^{-m}-1) [\log (s\tau)+\gamma_{\rm E} ]}{2 s^2}\label{asymptoticLaplace1} \nn \\
\fl & &  +2 \frac{e^{-m} [\log (m)+\gamma_{\rm E} ]-(m+1) \mbox{Ei}(-m)}{s} + \mathcal{O}\big(\ln(s)\big)\ , \\ 
\fl \tilde Z_{1AB}(m,s) &\stackrel{m \to \infty}{\simeq}&-\frac{\log (s \tau )+\gamma_{\rm E}}{2 s^2}+\mathcal{O}(e^{-m})\label{asymptoticLaplace2}\ ,
\end{eqnarray}
and
\begin{eqnarray}\label{asymptoticLaplace3}
\fl \int_0^{\infty}\!\!\!\rmd m \,e^m \partial_m \tilde{Z}_{1AB}^+(m,s)\stackrel{s \to 0}{\simeq}-\frac{\log (s \tau )+\gamma_{\rm E}-\frac{1}{2} }{ s^3}-\frac{\log (s \tau^3)+3 \gamma_{\rm E} +2}{ s^2} + \mathcal{O}\Big(\frac1s\Big)\ . \nn\\
\end{eqnarray}
Note that for the last term it is important to compute the integral over $m$   before expanding in $s$.

The remaining order-$\varepsilon $ correction in Eq.~\eqref{SurvivalPathIntegral} is due to a change of the diffusive constant in the Brownian action, from $D=1$ to $D_{\varepsilon ,\tau}=1+2\varepsilon (1+\ln\tau) +\mathcal{O}(\varepsilon )$, with the corresponding change in the drift such that the term linear in $z_t$ in $\mathcal{S}_0$, \textit{c.f.}\ Eq.~\eqref{Action}, remains unchanged. This change is equivalent to setting $T \to D_{\varepsilon ,\tau} T$ in the result for the Brownian, which, as stated in the main text, gives an order-$\varepsilon$ correction of the form \be
Z_{1D}(m,T)= 2(1+\ln \tau)T\partial_T Z_0^+(m,T)
\ee in Eq.~\eqref{expansion}, for a total first-order contribution \be
Z_1(m,T) = Z_{1AB}(m,T)+Z_{1D}(m,T)\ . 
\ee
The rescaling term $Z_1^D(m,T)$ contributes to the Pickands constant with  
\begin{equation}\label{B12}
 \int_0^{\infty}\!\!\!\rmd m \,e^m \partial_m {Z}_{1D}(m,T)\stackrel{T \to \infty}{\simeq} 2(1+\ln \tau) T + \mathcal{O}(e^{-T/4})\ .
\end{equation}
The inverse Laplace transform of Eq.~\eqref{asymptoticLaplace3} plus the contribution from (\ref{B12}) gives the result \eqref{Asymptotic1} of the main text. For the two other terms, the rescaling of the diffusive constant has no impact as
\begin{equation}
\lim\limits_{T\to \infty} T \partial_T Z_0^+(m,T)=\lim\limits_{m\to \infty} T \partial_T Z_0^+(m,T) = 0\ .
\end{equation}
Finally, formulae \eqref{Asymptotic2} and \eqref{AsymptoticNormalisation} are computed directly from \eqref{asymptoticLaplace1} and \eqref{asymptoticLaplace2} via an inverse Laplace transformation.
\section{Heuristic derivation  of the conjecture \eqref{conjecture}}
\label{A:conjecture}
The heurisitic derivation of our conjecture \eqref{conjecture} is as follows:
For $m \ll T^{\alpha}$ and $T \gg1$ we have $\mathcal{P}^T_{\alpha}(m)\simeq\mathcal{P}^{\infty}_{\alpha}(m)$, while for $m \gg T^{\alpha}$,   up to subleading (power-law and constant) corrections $\mathcal{P}^{T}_{\alpha}(m) \simeq e^{-\frac{(m-T^{\alpha})^2}{4 T^{\alpha}}}$, since very large values of the maximum are reached towards the end of the  time interval. Using that this cutoff function becomes sharp for large $T$,   we get
\begin{equation}
\frac1T \int_0^{\infty}\!\! \rmd m\, e^{m} \mathcal{P}_{\alpha}^T(m) \simeq \frac1T\int_0^{T^{\alpha}} \!\! \rmd m\, e^{m} \mathcal{P}^{\infty}_{\alpha}(m)  \ .
\end{equation}
In order to make Pickands' definition meaningful,  the r.h.s.\ has to become independent of $T$ for large $T$. This implies that the large-$m$ behaviour of $\mathcal{P}^{\infty}_{\alpha}(m)$ is exponentially decaying in $m$ to compensate the $e^m$ prefactor. This can still be multiplied by  a power law in $m$ times a constant.   The unique such possibility is 
\be
 \mathcal{P}^\infty_{\alpha}(m) \simeq  \frac{\mathcal{H}_{\alpha}}{\alpha} m^{\frac{1}{\alpha}-1} e^{-m}\ , 
\ee
as given in Eq.~(\ref{conjecture}).

\section{Extracting the Pickands constant from the maximum of a fBm bridge}
\label{app:bridge2Pickands}
Theorem (\ref{PiterbargTh})  applies to a  fractional Brownian bridge defined on $[0,1]$. Normalizing the process s.t.\ $\E(X_{t=1/2}^2)=1$,    Eqs.~(\ref{12})--(\ref{13}) are satisfied with 
\begin{equation}\label{PiterbargTh2}
    \alpha=2 H\;,\quad \beta=2\;, \quad  a=\frac{4 \alpha  (2^{1-\alpha } \alpha -\alpha +1)}{4-2^{\alpha
   }}\;, \quad  c=\frac{2^{\alpha +1}}{4-2^{\alpha }}\ .
\end{equation}
  Expanding Eq.~(\ref{PiterbargTh}) in $\alpha-1$ yields 
\begin{eqnarray}
&&\fl \partial_u  \mathbb{P} \left( \mbox{max}_{t\in[0,1]} X_t>u\right)  \simeq  {\cal H_{\alpha} } \,u\, \rme^{-\frac{u^{2}}2}\Big\{ 1+  \Big[ \ln (4) -4 \ln (u)-1 \Big]\frac{\alpha-1}2 + {\cal O}(\alpha-1)^{2}\Big\} \ . \nn\\
\end{eqnarray}
Our result (90) from Ref.~\cite{DelormeWiese2016b}, valid at order $\alpha-1$,  and  expanded for large $u$  is
\begin{eqnarray}
\fl && \partial_u \mathbb{P} \left( \mbox{max}_{t\in[0,1]} X_t>u\right) \nn \\
\fl &&~~ \simeq    u\, \rme^{-\frac{u^{2}}2}\Big\{ 1+  \Big[ \ln (4) -4 \ln (u)-1 -2\gamma_{{\rm E}}\Big]\frac{\alpha-1}2  +{\cal O}(\alpha-1)^{2} +{\cal O}(u^{-1})\Big\}  
\end{eqnarray}
This identifies
$
{\cal H_{\alpha} } = 1-\gamma_{{\rm E}} (\alpha-1)+{\cal O}(\alpha-1)^{2},
$
confirming Eq.~(\ref{us-Pickands}).

\begin{figure}
\centerline{\includegraphics[width=9cm]{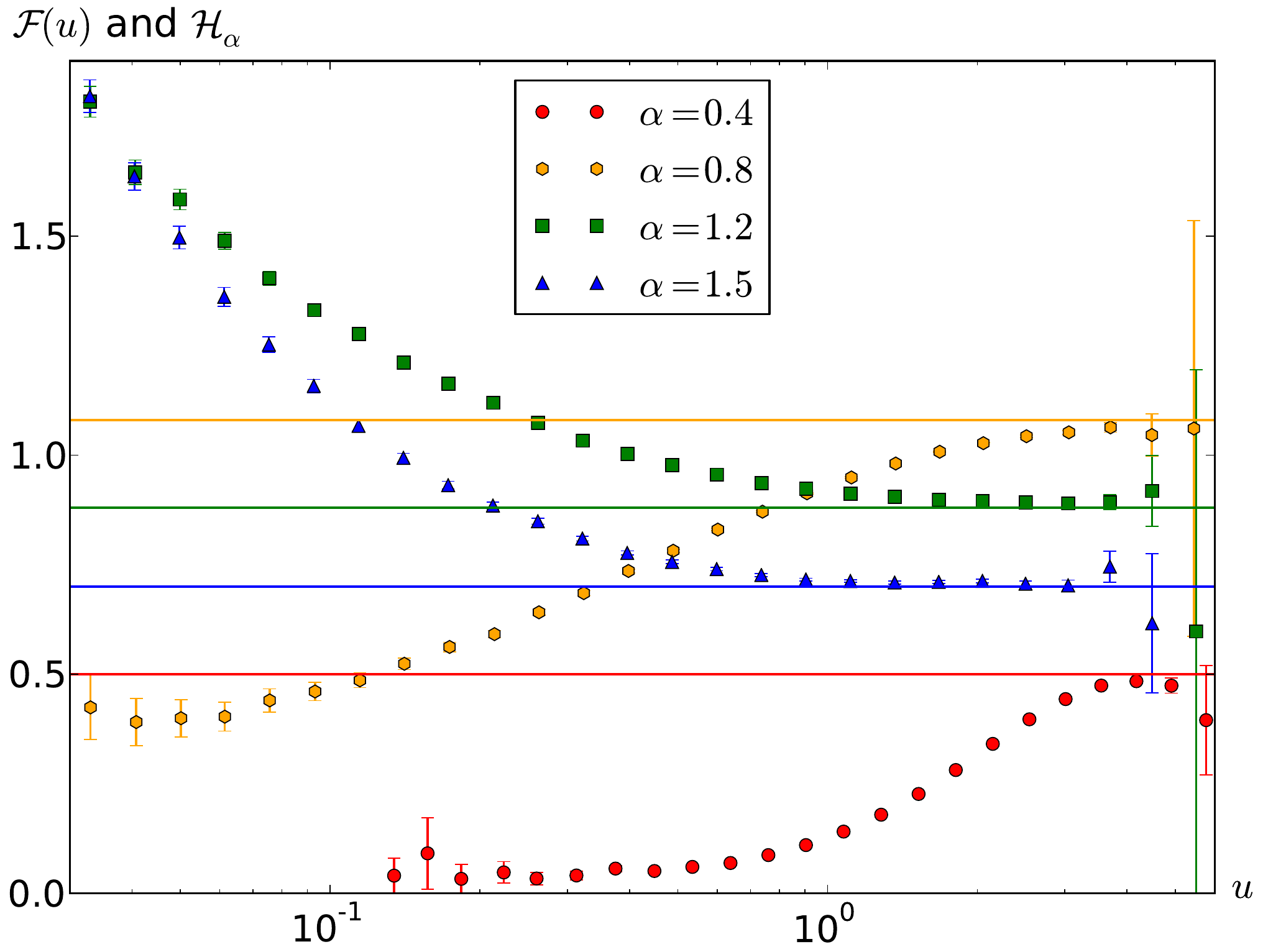}}
\caption{Plot of ${\cal F}(u):= \mathbb{P} \big(\max\limits_{t\in[0,1]} X_t=u\big)  u^{1-\frac{2}{\alpha} } \rme^{u^2/2} \sqrt{2 a}\, c^{-\frac{1}{\alpha}} $, and its convergence to ${\cal H}_\alpha$ for $u\to \infty$. The constants $a$ and $c$ are given in Eq.\ \eqref{PiterbargTh2}. The symbols are simulations for $\alpha=0.4$  to $\alpha=1.5$, cf.\ legend, with numerical parameters $T/dt=N=2^{18}$ and $10^6$ samples. Plain lines of the same color are the estimated asymptotics at large $u$, i.e. the Pickands constant, leading to the gray-blue results presented in Figure \ref{f:Dieker-data}.}
\end{figure}


\vspace{5mm}




\begin{thebibliography}{10}

\bibitem{Majumdar2010}
S.~N. {Majumdar},
\newblock \textit{{Universal first-passage properties of discrete-time random
  walks and {L{\'e}vy} flights on a line: Statistics of the global maximum and
  records}},
\newblock Physica A {\bf 389} (2010)   4299--4316.

\bibitem{BrayMajumdarSchehr2013}
A.~J. Bray, S.~N. Majumdar  and G. Schehr,
\newblock \textit{Persistence and first-passage properties in nonequilibrium
  systems},
\newblock Advances in Physics {\bf 62} (2013)   225--361.

\bibitem{DerridaSpohn1988}
B. Derrida and H. Spohn,
\newblock \textit{Polymers on disordered trees, spin glasses, and traveling
  waves},
\newblock J. Stat. Phys. {\bf 51} (1988)   817--40.

\bibitem{TracyWidom1994}
C.~A. {Tracy} and H.~{Widom},
\newblock \textit{Level-spacing distributions and the Airy kernel},
\newblock Comm. Math. Phys. {\bf 159} (1994)   151--174,
\newblock arXiv:hep-th/{\bf 9211141}.

\bibitem{Sire2007}
C. Sire,
\newblock \textit{Probability distribution of the maximum of a smooth temporal
  signal},
\newblock Phys. Rev. Lett. {\bf 98} (2007)   020601.

\bibitem{Sire2008}
C. Sire,
\newblock \textit{Crossing intervals of non-{Markovian Gaussian} processes},
\newblock Phys. Rev. E {\bf 78} (2008)   011121.

\bibitem{PiterbargBook1995}
V.~I. Piterbarg,
\newblock \textit{Asymptotic Methods in the Theory of Gaussian Processes and
  Fields},
\newblock Translations of Mathematical Monographs, vol.\ 148,
\newblock American Mathematical Society, 1995.

\bibitem{PiterbargBook2015}
V.~I. Piterbarg,
\newblock \textit{Twenty Lectures About {Gaussian} Processes},
\newblock Atlantic Financial Press, 2015.

\bibitem{Harper2014}
A.~J. Harper,
\newblock \textit{Pickands' constant {${\mathcal H}_\alpha$} does not equal
  {$1/\Gamma(1/\alpha)$}, for small $\alpha$},
\newblock arXiv:\null {\bf 1404.5505} (2014).

\bibitem{Michna2009}
Z. Michna,
\newblock \textit{Remarks on {Pickands} theorem},
\newblock arXiv:\null {\bf 0904.3832} (2009).

\bibitem{DebickiKisowski2008}
K. D{\c e}bicki and P. Kisowski,
\newblock \textit{A note on upper estimates for {Pickands} constants},
\newblock Statist. Probab. Lett. {\bf 78} (2008)   2046--2051.

\bibitem{HaanPickands1986}
L. de~Haan and J. Pickands,
\newblock \textit{Stationary min-stable processes},
\newblock Probability Theory and RelatedFields {\bf 72} (1986)   477--492.

\bibitem{Pickands1969}
{J. Pickands III},
\newblock \textit{Asymptotic properties of the maximum in a stationary
  {Gaussian} process},
\newblock Trans. Amer. Math. Soc. {\bf 145} (1969)  ~75.

\bibitem{Borell1976}
C. Borell,
\newblock \textit{The {Brunn-Minkowski} inequality in {Gauss} space},
\newblock Invent. Math. {\bf 30} (1976)   207--216.

\bibitem{MandelbrotVanNess1968}
B.~B. Mandelbrot and J.~W. {Van~Ness},
\newblock \textit{Fractional {B}rownian motions, fractional noises and
  applications},
\newblock SIAM Review {\bf 10} (1968)   422--437.

\bibitem{DiekerYakir2014}
A.~B. Dieker and B. Yakir,
\newblock \textit{On asymptotic constants in the theory of extremes for
  {Gaussian} processes},
\newblock Bernoulli {\bf 20} (2014)   1600--1619,
\newblock arXiv:1206.5840v3.

\bibitem{DelormeWiese2016b}
M. Delorme and K.~J. Wiese,
\newblock \textit{Extreme-value statistics of fractional {Brownian} motion
  bridges},
\newblock arXiv:\null {\bf 1605.04132} (2016);  Phys. Rev. E. (in print).

\bibitem{WieseMajumdarRosso2010}
K.~J. Wiese, S.~N. Majumdar  and A. Rosso,
\newblock \textit{Perturbation theory for fractional {Brownian} motion in
  presence of absorbing boundaries},
\newblock Phys. Rev. E {\bf 83} (2011)   061141,
\newblock arXiv:{\bf 1011.4807}.

\bibitem{DelormeWiese2015}
M. Delorme and {K.~J.} Wiese,
\newblock \textit{The maximum of a fractional {Brownian} motion: Analytic
  results from perturbation theory},
\newblock Phys. Rev. Lett. {\bf 115} (2015)   210601,
\newblock arXiv:{\bf 1507.06238}.

\bibitem{DelormeWiese2016}
M. Delorme and K.~J. Wiese,
\newblock \textit{Perturbative expansion for the maximum of fractional
  {Brownian} motion},
\newblock Phys. Rev. E {\bf 94} (2016)   012134,
\newblock arXiv:{\bf 1603.00651}.

\bibitem{MajumdarSire1996}
S.~N. Majumdar and C. Sire,
\newblock \textit{Survival probability of a {G}aussian non-{M}arkovian process:
  Application to the
  $\mathit{T}\phantom{\rule{0ex}{0ex}}=\phantom{\rule{0ex}{0ex}}0$ dynamics of
  the {I}sing model},
\newblock Phys. Rev. Lett. {\bf 77} (1996)   1420--1423.

\bibitem{SireMajumdarRudinger2000}
C. Sire, S.~N. Majumdar  and A. R\"udinger,
\newblock \textit{Analytical results for random walk persistence},
\newblock Phys. Rev. E {\bf 61} (2000)   1258--1269.

\bibitem{DelormeWieseUnPublished}
M. Delorme and {K.~J.} Wiese,
\newblock unpublished.

\end{thebibliography}
\end{document}